\documentclass[
reprint,
superscriptaddress,
 amsmath,amssymb,
aps,
]{revtex4-2}

\usepackage{color}
\usepackage{graphicx,physics,mathtools}% Include figure files
\usepackage{dcolumn}% Align table columns on decimal point
\usepackage{bm}% bold math
\usepackage{float}
\usepackage[breaklinks,colorlinks=true,linkcolor=blue,urlcolor=blue,citecolor=blue]{hyperref}
\usepackage[normalem]{ulem}

\usepackage{graphicx}
\usepackage{epstopdf}

\begin{document}

%\preprint{APS/123-QED}

\title{Multi-parameter estimation enhanced by level crossing Hamiltonian structure with Bayesian feedback}% Force line breaks with \\
\author{Ryo Tatsumi}
 \email{tatsumi.ryo.p6@dc.tohoku.ac.jp}
 \affiliation{Department of Applied Physics, Graduate School of Engineering, Tohoku University, Sendai, Miyagi 980-8579, Japan}%Lines break automatically or can be forced with \\
\author{Le Bin Ho}
 \email{ho.bin.le.e3@tohoku.ac.jp}
 \affiliation{Department of Applied Physics, Graduate School of Engineering, Tohoku University, Sendai, Miyagi 980-8579, Japan}
 \affiliation{Frontier Research Institute for Interdisciplinary Sciences, Tohoku University, Sendai, Miyagi 980-8578, Japan}
\author{Hiroaki Matsueda}
 \affiliation{Department of Applied Physics, Graduate School of Engineering, Tohoku University, Sendai, Miyagi 980-8579, Japan}
 \affiliation{Center for Science and Innovation in Spintronics, Tohoku University, Sendai 980-8577, Japan}

%\affiliation{}%

\date{\today}% It is always \today, today,
 
\begin{abstract}
%------------------------------
%In the context of quantum metrology, which enables high-precision parameter estimation using quantum states, Quantum Fisher Information (QFI) serves as the sensitivity of an estimator to the state and is a crucial concept. In this study, we analyzed a general two-level system with level repulsion to confirm the universal behavior of QFI in quantum systems. For multi-parameter estimation using practical quenching processes, we demonstrated the conditions under which QFI for multiple parameters is simultaneously maximized at the level crossing point. Furthermore, by introducing Bayesian estimation methods, we explicitly constructed and numerically verified a resource-efficient estimation process achieving high-precision estimation. This estimation process is universally applicable to any quantum system exhibiting a level repulsion structure. This work provides new guidance for high-sensitivity, low-cost multi-parameter estimation process across diverse quantum platforms, thereby expanding fundamental knowledge for applied quantum metrology.
Level crossings, where two quantum energy levels intersect, are ubiquitous in quantum systems and play a central role in controlling quantum dynamics.
In this study, we establish a framework that connects these level crossing structures to the ultimate precision limits of quantum metrology.
By analyzing a Hamiltonian with the level crossings structure, we identify the conditions under which multiple parameters can be estimated simultaneously with maximal sensitivity near the level crossing point. 
We then construct a resource-efficient sensing protocol based on unitary quench dynamics and Bayesian estimation, enabling high-precision, low-cost parameter estimation without requiring entanglement or complex control.
We present a protocol whose estimation precision saturates the Holevo Cram\'er–Rao bound, the tightest achievable precision bound in multi-parameter estimation.
Using the two-level model and the two-spin Ising model as illustrative examples, we demonstrate how the protocol can be applied to a system with level crossing Hamiltonian structure.
%The validity of this protocol is confirmed through both analytical derivations and numerical simulations, including an application to the two-spin Ising model.
Our findings reveal that level crossings serve as universal operating points for precision quantum metrology, offering a simple and scalable path toward practical multi-parameter sensing across diverse quantum platforms.
%------------------------------
\end{abstract}

%\pacs{}

\maketitle
\section{Introduction}
Quantum metrology, which enables high-precision parameter estimation using quantum mechanics, is one of the central research themes in quantum information science and precision measurement \cite{Giovannetti04,Giovannetti06}.
Quantum Fisher information (QFI) \cite{Paris04,Pezze14} sets the ultimate precision bound through the quantum Cramér--Rao (CR) inequality and quantifies the sensitivity of a quantum state to small variations in the parameter of interest. 
Consequently, the properties of QFI have been investigated theoretically and experimentally in various quantum systems \cite{Giovannetti06,Zanardi07,Yu21}.
Numerous approaches have been proposed to improve measurement precision by increasing QFI, particularly through the use of quantum entanglement and squeezing \cite{Pezz09,Toth12,Hyllus12,Ma11,Pezze18,PhysRevA.102.022602,PhysRevResearch.6.033292}.

Recently, attention has shifted from resource-based enhancement toward exploiting intrinsic Hamiltonian properties that enhance parameter sensitivity. A prominent example is quantum critical metrology, which uses ground states near quantum phase transitions as sensitive probes \cite{Frerot18,Mihailescu25}. 
Near a critical point, small parameter variations can induce pronounced changes in the ground state, leading to enhanced sensitivity. 
However, preparing a quantum state near criticality requires increasingly long adiabatic preparation times and increases vulnerability to decoherence, posing a major challenge for practical implementation.

A closely related and more general phenomenon is the appearance of level crossings or avoided crossings (anticrossings), where the level crossing is avoided by perturbation.
Near an anticrossing, the ground state can vary rapidly with the parameters, resulting in a significant enhancement of the QFI \cite{Ghirardi18,Mihailescu25}. 
However, previous studies have largely focused on static ground-state sensing and single-parameter estimation, while realistic sensing applications often require the simultaneous estimation of multiple noncommuting parameters.

Multi-parameter quantum metrology introduces fundamental challenges absent in single-parameter estimation \cite{Zhu15,Szczykulska16}. When the optimal measurements for different parameters are incompatible, the corresponding quantum CR bound generally cannot be saturated simultaneously \cite{Ragy16}. A tighter attainable precision limit is provided by the Holevo CR bound, which explicitly accounts for such measurement incompatibility \cite{Sidhu21,Suzuki16,Holevo11,Hayashi08,Albarelli19}. Developing experimentally feasible protocols that approach this bound without requiring complex entangled resources remains a central challenge in quantum sensing.

%In this work, we first analyze a two-level system to investigate how the estimation performance evolves in multi-parameter estimation based on unitary time evolution for quantum systems with level-crossing structures. We confirm that by carefully selecting the initial state, the QFI for all estimated parameters simultaneously achieves its maximum values at the level-crossing point. Furthermore, by introducing the classical Fisher information and an adaptive estimation strategy \cite{Montenegro21,Salvia23}, we construct a concrete measurement process that achieves high-precision estimation using the chosen measurement operators. By deriving the Holevo Cramér–Rao bound at the crossing point, we show that the selected measurement operators are optimal among all possible measurement operators. This process employs quench dynamics, which requires no time-dependent external field tuning and avoids using entangled states, enabling the completion of the protocol with few resources. Consequently, it is expected to find future applications as a low-cost quantum sensor for edge devices. Furthermore, this method is applicable to quantum systems exhibiting a level repulsion structure. 

In this work, we investigate multi-parameter estimation based on unitary dynamics in quantum systems exhibiting level crossings. Starting with a two-level system, we show that an appropriate choice of the initial state allows the QFI for all parameters to simultaneously reach their maximum values at the crossing point. We then construct an adaptive measurement protocol based on the classical Fisher information \cite{Montenegro21,Salvia23} and show, by deriving the Holevo CR bound, that the proposed measurements are optimal at the crossing point. The protocol relies on quench dynamics and requires neither time-dependent control of external fields nor entangled probe states, thereby reducing the resources required for implementation. %Finally, we show that the approach extends to systems exhibiting level repulsion, demonstrating its applicability beyond exact level crossings.

%Indeed, using the two-spin Ising model as an example, we verified the measurement protocol for multi-parameter estimation and the conditions under which the protocol works well through numerical simulations. \blue{To investigate the influence of the other energy levels, we varied the exchange interaction strength, thereby changing the energy separation between the crossing levels and the remaining levels. As the exchange interaction was increased and the crossing levels became more energetically isolated from the other levels, the behavior of the QFI was better described by the corresponding two-level model. However, energy levels other than the crossing levels also contribute to parameter estimation by facilitating global sensing.} It presents a comprehensive sensing scheme that covers the identification of the fundamental quantum precision limit, the construction and optimization of a practical measurement, and the numerical implementation of global parameter estimation. Because level crossings and effective two-level structures occur widely in spin systems, magnetic materials, quantum dots, superconducting circuits, and other quantum platforms, our results provide a broadly applicable foundation for simple, scalable, and resource-efficient multi-parameter quantum sensors.

We further validate the protocol using a two-spin Ising model and identify the regime in which the effective two-level description remains accurate. By varying the exchange interaction, we control the energy separation between the crossing levels and the remaining spectrum. As the crossing levels become increasingly isolated, the QFI approaches the behavior predicted by the corresponding two-level model. Away from this regime, higher-energy levels also contribute to the estimation, enabling sensitivity over a broader parameter range. 

Together, these results establish a complete sensing framework that connects the fundamental precision bound, the construction of an optimal measurement strategy, and its implementation for global parameter estimation. Since level crossings and effective two-level structures arise naturally in a wide range of quantum systems, our approach provides a general framework for resource-efficient multi-parameter sensing.

The remainder of this paper is organized as follows. 
In Sec.~\ref{Sec:model}, we develop the multi-parameter sensing framework for a two-level system, including the optimal measurement strategy and Bayesian adaptive estimation. 
In Sec.~\ref{Sec:Ising}, we extend the framework to a two-spin Ising model and examine the role and validity of the effective two-level description. 
Finally, Sec.~\ref{Sec:Conclusion} summarizes our conclusions.

\section{Two level system}\label{Sec:model}

%-----------------------figure 1------------------------%
\begin{figure}[t]
\centering
\includegraphics[width=1.0\columnwidth]{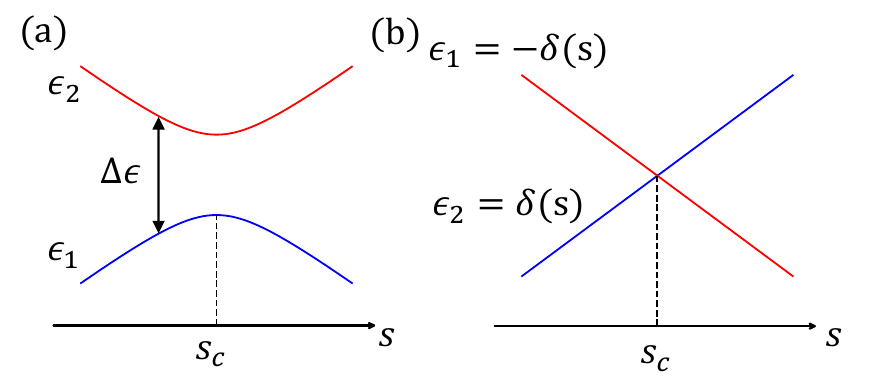}
\caption{The eigenvalue of the system's Hamiltonian  as a function of the parameter $s$ where (a) $q(a) \neq 0$, (b) $q(a) = 0$.}
\label{fig:Eigenvalue}
\end{figure}
%-----------------------figure 1------------------------%

\subsection{Model}
%Let us consider a simple model of two-level system governed by a parameter-depending Hamiltonian
We consider a two-level system described by a Hamiltonian that depends on two parameters, \(s\) and \(a\):
\begin{equation} \label{Eq:general_Hamiltonian}
\begin{split}
        H = \begin{pmatrix}
        \delta(s) & q(a)\\
        q(a) & -\delta(s)
    \end{pmatrix}
    =  \delta(s) \sigma_z + q(a) \sigma_x,
\end{split}
\end{equation}
%where $s$ and $a$ are estimated parameters, $\delta(s)$ and $q(a)$ are $s$ and $a$-dependent parameters.
%Physically, $\delta(s)$ represents a tunable external field that controls the energy splitting between the states, while $q(a)$ characterizes a transverse field that opens a gap at the avoided crossing. 
%This model is well-suited for quantum metrology, where estimating $s$ and $a$ enables precise sensing of external fields.
Here, $\delta(s)$ represents a longitudinal field that controls the energy splitting between the two levels, while $q(a)$ is a transverse field that couples the levels and opens a gap at an avoided crossing. 
The goal is to estimate both parameters $s$ and $a$, making this model relevant for quantum sensing.

The energies $\epsilon_{1(2)}$ and the gap $\Delta\epsilon$ are given by 
\begin{equation}
\begin{split}
    \epsilon_{1} &= -\sqrt{\delta(s)^2 + q(a)^2}, \\
    \epsilon_{2} &= \sqrt{\delta(s)^2 + q(a)^2},\\
    \Delta\epsilon &= 2 \sqrt{\delta(s)^2 + q(a)^2}.
\end{split}
\end{equation}
%\blue{At the critical value $s = s_c$, where $\delta(s_c) = 0$, the two energy levels are closest to each other as $s$ is varied.} 
A critical point occurs at \(s = s_c\), defined by $\delta(s_c) = 0$, where the two energy levels approach each other most closely. Similarly, we define \(a_c\) by \(q(a_c) = 0\).
%Similarly, we define the value $a_c$ as satisfying $q(a_c) = 0$.
When $a \neq a_c$, i.e., $q(a) \neq 0$, the energy gap remains finite at $s = s_c$, indicating an avoided crossing [Fig.~\ref{fig:Eigenvalue}(a)].
In contrast, when $a = a_c$, the energies reduce to $\epsilon_{1(2)} = \mp \delta(s)$, and the gap $\Delta \epsilon = 2|\delta(s)|$ vanishes at $s=s_c$, as shown in Fig.~\ref{fig:Eigenvalue}(b). We refer to the point $(s_c, a_c)$ as the crossing point.

The corresponding eigenstates of the Hamiltonian are given by 
\begin{equation}\label{Eq:eigen_state}
\begin{split}
    \ket{\eta_1} = \sin\left(\frac{\phi}{2} \right) \ket{0} - \cos\left(\frac{\phi}{2}\right)\ket{1},\\
    \ket{\eta_2} = \cos\left(\frac{\phi}{2}\right) \ket{0} +\sin\left(\frac{\phi}{2}\right)\ket{1},
\end{split}
\end{equation}
where $\tan{\phi} = q(a)/\delta(s)$, $\ket{0} = (1,0)^\top$ and $\ket{1} = (0,1)^\top$.

%\section{Quantum Fisher information}
\subsection{QFI with ground state}

%-----------------------figure 2-------------------------
\begin{figure*}[t]
\centering
\includegraphics[width=0.7\linewidth]{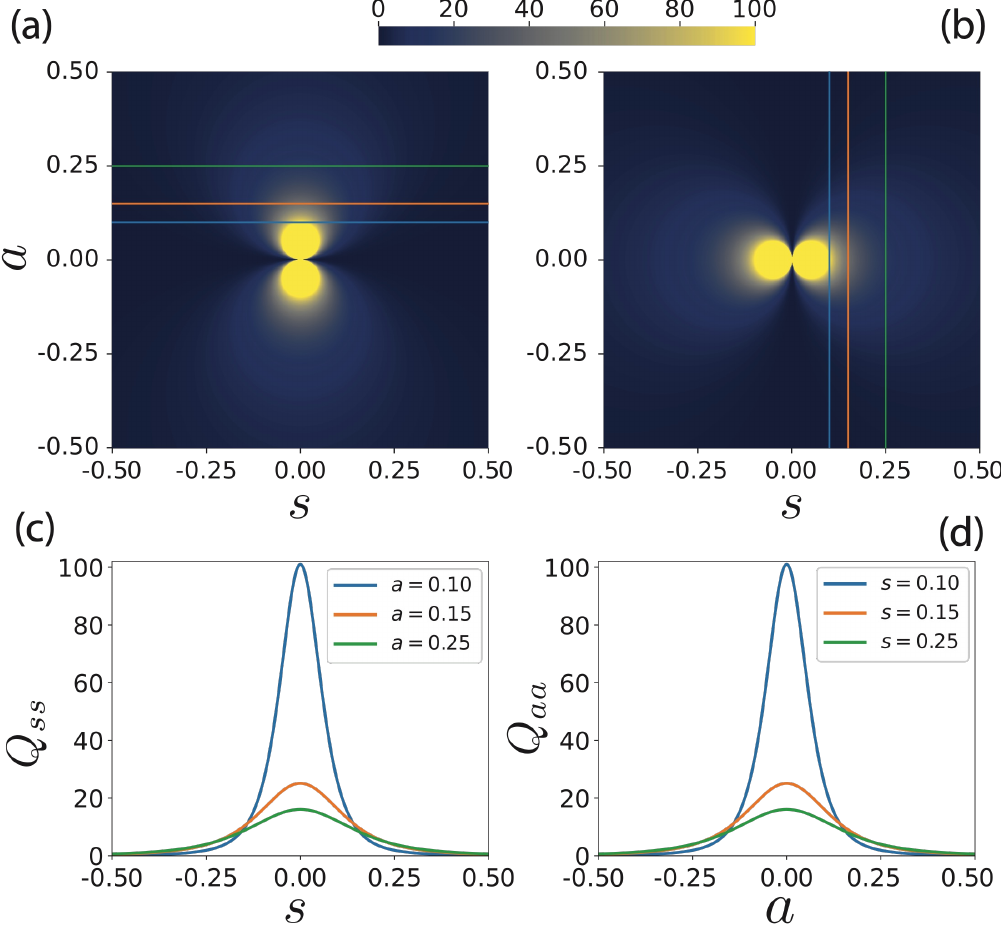}
\caption{\textbf{QFI for the ground state of a two-level system.}
Heatmaps of (a) $Q_{ss}$ and (b) $Q_{aa}$ computed from the ground state.
(c) $Q_{ss}$ as a function of $s$ for several fixed values of $a$, and (d) $Q_{aa}$ as a function of $a$ for several fixed values of $s$. In all panels, the QFI increases as the system approaches the crossing point $(s,a)=(0,0)$, where it becomes sharply enhanced.}
\label{fig:QFI_ground}
\end{figure*}
%-----------------------------------------------------

%-----------------------figure 3-------------------------
\begin{figure*}[t]
\centering
\includegraphics[width=2.05\columnwidth]{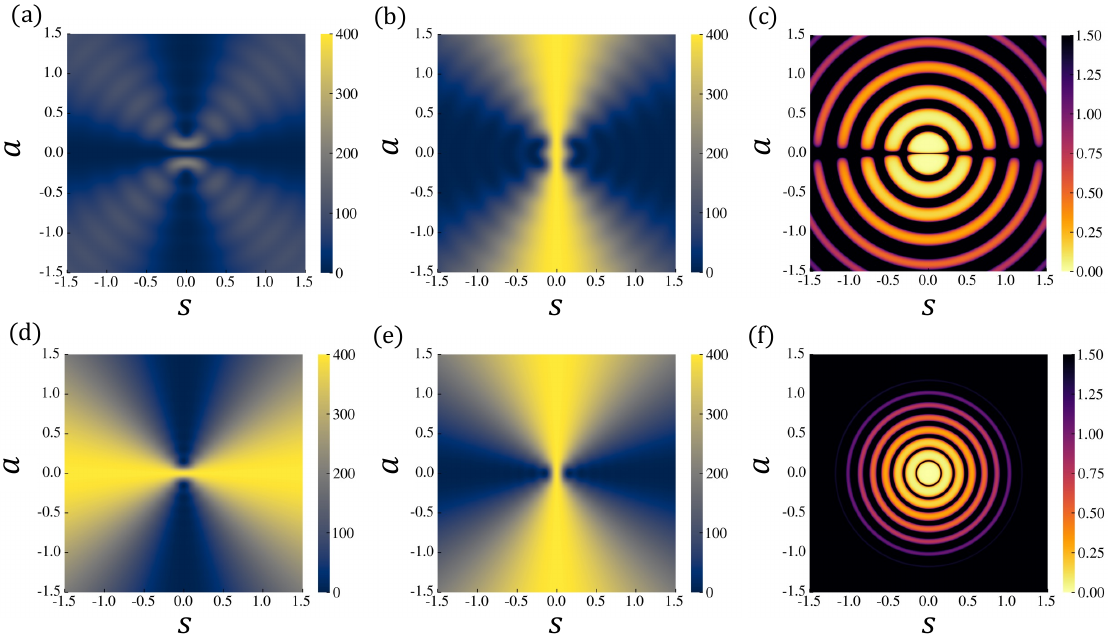}
\caption{\textbf{QFI with a unitary process in two-level systems.}
The heatmaps of (a) $Q_{ss}$, (b) $Q_{aa}$, and (c) $\mathrm{Tr}[\bm{\mathrm{Q}}^{-1}]$ with initial state has $\theta = 0$.
The heatmaps of (d)$Q_{ss}$, (e)$Q_{aa}$, and (f)$\mathrm{Tr}[\bm{\mathrm{Q}}^{-1}]$ with initial state has $\theta = \pi/2$.}
\label{fig:QFI_2_z}
\end{figure*}
%-----------------------------------------------------

%Now, we estimate the parameters $s$ and $a$ using the ground state of the system. We use the quantum Fisher information matrix (QFIM) and its relation to the variance of physical quantities to evaluate the metrological gain of the estimation process. For a pure quantum state $\ket{\psi}$, the QFIM is defined as
We estimate the parameters $s$ and $a$ using the ground state of the system and quantify the precision via the quantum Fisher information matrix (QFIM). For a pure state $|\psi\rangle$, the QFIM is defined as
\begin{align}
\mathbf{Q} = 4 \mathrm{Re} \left[ \mathbf{J} - \mathbf{v} \mathbf{v}^\dagger \right],
\end{align}
where $\mathbf{J}$ is the Jacobian inner-product matrix with elements
$J_{\nu\mu} = \langle \partial_{\mu} \psi | \partial_{\nu} \psi \rangle$, and
$\mathbf{v}$ is a complex vector with elements $v_\mu = \langle \psi | \partial_{\mu} \psi \rangle$, for $\{\mu, \nu\} \in \{s, a\}$.
%The total estimation variance satisfies ${\rm var}(s) + {\rm var}(a) \geq \mathrm{Tr}[\mathbf{Q}^{-1}]/n$, where $n$ is the number of repeated measurements.
The total estimation error is bounded by ${\rm var}(s) + {\rm var}(a) \ge \mathrm{Tr}[\mathbf{Q}^{-1}]/n$, where $n$ is the number of measurements.

For $a = a_c$, the ground state is $\ket{\eta_1} = \ket{0}$ for $s < s_c$, and switches to $\ket{\eta_2}  = \ket{1}$ for $s > s_c$.
Thus, $Q_{ss}$ is zero in these regions. 
At $s = s_c$, however, the ground state changes abruptly and discontinuously.
Therefore, $Q_{ss} \sim \delta_{\rm D}(s - s_c)$, resembling a Dirac delta function.

For $a \neq a_c$, the energy gap at $s=s_c$ remains finite, and the ground state varies smoothly with the parameters. In this regime, the QFIM takes the form
\begin{align}\label{Eq:QFIM_ground}
\mathbf{Q} = \frac{1}{(\delta^2 + q^2)^2} \begin{pmatrix}
q^2 (\partial_s \delta)^2 & q \delta(\partial_s \delta) (\partial_a q) \\
q \delta (\partial_s \delta) (\partial_a q) & \delta^2 (\partial_a q)^2
\end{pmatrix}.
\end{align}
However, $\det \mathbf{Q} = 0$ for all $\delta$ and $q$, implying that $\mathbf{Q}$ is not invertible. This demonstrates that the ground state does not provide sufficient information for simultaneous estimation of both parameters \cite{Liu20}.
%The inverse of the QFIM $\bm{\mathrm{Q}}^{-1}$ in Eq.~(\ref{Eq:QFIM_ground}), does not exist for any $\delta$ and $q$, since $\det \bm{\mathrm{Q}} = 0$.This indicates that, for multiparameter estimation, the ground state is not useful.

To illustrate these features, we consider the minimal model
\begin{equation}
H = \begin{pmatrix}
s & a\\
a & -s
\end{pmatrix}.
\label{simple_two}
\end{equation}
In this case, $\delta(s) = s$, $q(a) = a$, %and $\partial_s \delta(s) = \partial_a q(a) = 1$.
so the crossing point is $(s_c, a_c) = (0,0)$.
%We note that many level-repulsion systems can be approximated by the Hamiltonian in Eq.~(\ref{simple_two}) near the crossing point.
%With Eq.~\ref{simple_two}, the QFIM becomes
This model captures the universal behavior near the crossing point in many physical systems.
Using Eq.~\eqref{simple_two}, the QFIM~\eqref{Eq:QFIM_ground} becomes
\begin{align}\label{Eq:QFIM_ground_simple}
\mathbf{Q} = \frac{1}{(s^2 + a^2)^2} \begin{pmatrix}
a^2 & sa \\
sa  & s^2
\end{pmatrix}.
\end{align}
The diagonal elements exhibit sharp peaks at the crossing point: $Q_{ss}$ diverges as $a$ approaches zero with $s$ fixed at zero, whereas $Q_{aa}$ diverges as $s$ approaches zero with $a$ fixed at zero.
Figures~\ref{fig:QFI_ground}(a) and (b) show the heatmap of the diagonal elements of QFIM, $Q_{ss}$ and $Q_{aa}$.
An upper cutoff is imposed on the color scale of the heat maps.
Both $Q_{ss}$ and $Q_{aa}$ exhibit two bright, circle-like regions.
Figure~\ref{fig:QFI_ground}(c) shows $Q_{ss}$ as a function of $s$ for several fixed values of $a$.
As same as Eq.~\eqref{Eq:QFIM_ground_simple}, $Q_{ss}$ attains its maximum at $s=0$.
Similarly, Fig.~\ref{fig:QFI_ground}(d) shows that $Q_{aa}$ is maximized at $a=0$ for each fixed value of $s$.
These results indicate that the sensitivity for each single parameter can be significantly enhanced by tuning the system close to the crossing point.
Nevertheless, since the QFIM remains a singular matrix for all parameter values, the ground state is insufficient for multi-parameter estimation. This limitation motivates the use of dynamical approaches, which we investigate next.

%\blue{Changing parameter $s$, the element $Q_{ss}$ attains its maximum in $s=0$ for any $a$ and diverges at the crossing point, as shown in Fig.~\ref{fig:QFI_ground}(a) and (c).
%Similarly, $Q_{aa}$ peaks at $a = 0$ and diverges at the crossing point, as shown in Fig.~\ref{fig:QFI_ground}(b) and (d).}
%These features indicate that tuning the system near the crossing point can significantly enhance the sensitivity.
%For global \blue{and single} parameter estimation, sensitivity can be enhanced by tuning the system close to the crossing point \cite{Ghirardi18}, as discussed later.
%On the other hand, the multiparameter estimation is prohibits for any parameters, because of the invertibility of the QFIM.
%Therefore, we next focus on quantum metrology based on a unitary process.

\subsection{QFI under unitary process with general two-level Hamiltonian}
\label{Sec:QFI_two_level}
%We consider a practical scenario in which the system is initialized in a parameter-independent pure state $\ket{\psi_i}$ and evolves unitarily under the Hamiltonian $H$. The time-evolution operator is given by
We consider a practical setting in which the system is initialized in a parameter-independent pure state $|\psi_i\rangle$ and evolves under the Hamiltonian Eq.~\eqref{Eq:general_Hamiltonian} for a time $t$. The corresponding unitary evolution is
\begin{equation}\label{eq:Unitary_gene}
U = e^{-iHt} = r_I \sigma_I + i \bm{r} \cdot \bm{\sigma},
\end{equation}
where $r_I = \cos(\Omega t)$ with $\Omega = \sqrt{\delta^2 + q^2}$, and $\bm{r} = (r_x, r_y, r_z) = -\frac{\sin(\Omega t)}{\Omega} (q, 0, \delta)$.
Although the QFI can be obtained analytically for arbitrary values of $s$ and $a$, the resulting expression is too complicated. We therefore present the analytical result only at $s = s_c$.
Then, we have [see Appendix \ref{App:QFI_Unitary}]
\begin{equation}\label{Eq:QFI_two_2pi}
\begin{split}
        \notag Q_{ss} &= 4 t^2 {\rm sinc}^2(q t)\big(1 - f^2\big) (\partial_s\delta)^2,\\
    \notag Q_{aa} &= 4t^2\big(1 - \ev{\sigma_x}^2\big) (\partial_aq)^2,\\
    \notag Q_{sa} &=Q_{as} = -4t^2f {\rm sinc}(qt)\ev{\sigma_x} \partial_s\delta \partial_aq,
\end{split}
\end{equation}
where $f = \sin(q t) \ev{\sigma_y} + \cos(q t) \ev{\sigma_z}$ with  
$\ev{\sigma} \equiv \langle\psi_i|\sigma|\psi_i\rangle$, and $\mathrm{sinc}(q t) = \sin(q t)/q t$.
Then, we have
\begin{align}
    \mathrm{Tr}[\bm{\mathrm{Q}}^{-1}] = \frac{\mathrm{sinc}^2(q t) \big(1 - f^2\big)(\partial_s\delta)^2 + (1-\ev{\sigma_x}^2) (\partial_aq)^2}{4t^2 \mathrm{sinc}^2(q t) \big(1-f^2-\ev{\sigma_x}^2 \big)(\partial_s\delta)^2( \partial_aq)^2}.
\end{align}

A particularly important case arises when $\langle \sigma_x \rangle = 0$, for which the off-diagonal elements vanish, $Q_{as} = Q_{sa} = 0$, and $\mathrm{Tr}[\bm{\mathrm{Q}}^{-1}]$ become maximized.
We therefore restrict to initial states of the form
\begin{align}\label{Eq:initial_state_two}
\ket{\psi_i} = \cos\left(\frac{\theta}{2}\right)\ket{0} + i \sin\left(\frac{\theta}{2}\right)\ket{1}, \quad \theta \in [0, \pi],
\end{align}
which satisfy $\langle \sigma_x \rangle = 0$, $\langle \sigma_y \rangle = \sin\theta$, and $\langle \sigma_z \rangle = \cos\theta$.
For this initial state, the QFIM is simplified, and in particular,
\begin{align}
    Q_{aa} = 4t^2 (\partial_a q)^2,
\end{align}
which is independent of the choice of $\theta$.
In this form, we also have 
\begin{equation}
Q_{ss} =
\begin{cases}
4 t^2 \mathrm{sinc}(q t)\sin^2(q t)(\partial_s \delta)^2,
& \text{ for } \theta = 0,\\
4 t^2 \mathrm{sinc}(q t)\cos^2(q t)(\partial_s \delta)^2,
& \text{ for } \theta = \pi/2.\nonumber
\end{cases}
\end{equation}
In particular, $\mathrm{sinc}(q t)\cos^2(q t)$ attains its maximum value at $q = 0$. 
However, since $Q_{ss}$ also depends on the factor $4t^2(\partial_s \delta)^2$, the value of $Q_{ss}$ at $\theta = \pi/2$ is not necessarily maximized at $\delta = q = 0$.

%$Q_{ss} = 4t^2{\rm sinc}(q t)\sin^2(q t)(\partial_s\delta)^2$ for $\langle \sigma_y \rangle = 0$, i.e., $\theta = 0; \ket{\psi_i} = \ket{0}$, and $Q_{ss} = 4t^2{\rm sinc}(q t)\cos^2(q t)(\partial_s\delta)^2$ for $\langle \sigma_z \rangle = 0$, i.e., $\theta = \pi/2; \ket{\psi_i} = (\ket{0}+i\ket{1})/\sqrt{2}$.

\subsection{Numerical and theoretical calculation of QFI with minimal two-level Hamiltonian}
We found that the QFI for Eq.~(\ref{Eq:general_Hamiltonian}) depends on derivative terms such as $\partial_s\delta$. 
To clarify the role of the level crossing, and to numerically examine the behavior away from $s\neq s_c$, we therefore focus on the minimal level-crossing model given in Eq.~\eqref{simple_two}.

Figures~\ref{fig:QFI_2_z}(a)–(c) show the numerical results for $Q_{ss}$, $Q_{aa}$, and $\mathrm{Tr}[\bm{\mathrm{Q}}^{-1}]$ under unitary evolution at time $t = 10$ with the initial state for $\theta = 0$ in Eq.~(\ref{Eq:initial_state_two}), using the Hamiltonian in Eq.~\eqref{simple_two}.
Both $Q_{ss}$ and $Q_{aa}$ peak at $s = 0$. 
However, whereas $Q_{aa}$ attains the maximum value $4t^2 = 400$ for all $a$, $Q_{ss}$ remains below $4t^2$ even at its maximum.
To confirm this behavior, we show analytical expressions for $Q_{ss}$ and $Q_{aa}$ at $s = 0$.
\begin{equation}
\begin{split}
    {Q}_{ss} &= 4 t^2 {\rm sinc}^2(a t) \sin^2(a t),\\
    {Q}_{aa} &= 4 t^2 ,\\
\end{split}
\end{equation}
where $Q_{ss}$ converges to 0 at $a = 0$ and attains its maximum at $a = a_0$, where $a_0 \simeq 1.16556/t$ as shown in Fig.~\ref{fig:QFI_2_z}(a).

As discussed below, $4t^2$ is the maximum possible value of $Q_{aa}$ for any initial state.
At $s = 0$, the unitary operator is given by $U = e^{-i a t \sigma_x}$, and therefore $Q_{aa}$ can be written as
\begin{equation}
Q_{aa} = 4t^2 \mathrm{var} (\sigma_x),
\label{var_H}
\end{equation}
where $\mathrm{var}(A) = \bra{\psi_i} A^2 \ket{\psi_i} - \bra{\psi_i} A \ket{\psi_i}^2$ \cite{Giovannetti06}.
In general, $\mathrm{var}(A)$ attains its maximum value when $\ket{\psi_i}$ is a superposition of the eigenstates corresponding to the maximum and minimum eigenvalues of $A$.
The maximum value of $\mathrm{var}(\sigma_x)$ is 1, which is attained when $\theta = 0$, i.e., $\ket{\psi_i} = \ket{0}$. 
This state can be written as a superposition of the eigenstates of $\sigma_x$ with maximum and minimum eigenvalues, $\ket{+} = (\ket{0} + \ket{1})/\sqrt{2}$ and $\ket{-} =(\ket{0} - \ket{1})/\sqrt{2}$.
Therefore, $4t^2$ is the maximum value of $Q_{aa}$ for any initial state at $s = 0$.
As shown in Fig.~\ref{fig:QFI_2_z}(c), $\mathrm{Tr}[\mathbf{Q}^{-1}]$ takes its minimum value at $(s,a) = (0,a_0)$ as
\begin{equation}
\begin{split}
\mathrm{Tr}[\mathbf{Q}^{-1}] = \frac{1}{4t^2} \left(1 + \frac{1}{\sin^2(2a_0 t)} \right).
\end{split}
\label{TrQ_sigmaz}
\end{equation}

We now switch to an initial state with $\theta = \pi/2$.
Figures~\ref{fig:QFI_2_z}(d)–(f) show the numerical results for $Q_{ss}$, $Q_{aa}$, and $\mathrm{Tr}[\bm{\mathrm{Q}}^{-1}]$ with $\theta = \pi/2$, using the Hamiltonian in Eq.~\eqref{simple_two}.
$Q_{ss}$ appears to attain its maximum along $a=0$, whereas $Q_{aa}$ is maximized along $s=0$. 
Consequently, $\mathrm{Tr}[\mathbf{Q}^{-1}]$ reaches its minimum at $(s,a)=(0,0)$, identifying the level-crossing point as a hot spot for multi-parameter estimation.
Furthermore, the heatmap shown in Fig.~\ref{fig:QFI_2_z}(d) is related to that in Fig.~\ref{fig:QFI_2_z}(e) by a $90^\circ$ rotation.
This relation originates from the symmetry of the Hamiltonian and the initial state under the interchange $s\leftrightarrow a$, which merely exchanges the labels of the two parameters.
In this case, analytical result of $Q_{ss}$ and $Q_{aa}$ for $s=0$ become
\begin{equation}
\begin{split}
Q_{ss} &= 4 t^2 {\rm sinc}^2(at)\cos^2(a t),\\
Q_{aa} &= 4 t^2.
\end{split}
\label{Eq:Qss_linear}
\end{equation}
$Q_{ss}$ attains its maximum value, $4t^2$, at $a = 0$.
From Eq.~\eqref{Eq:eigen_state}, the eigenstates at $a = 0$ are $\ket{\eta_1} = \ket{1}$ and $\ket{\eta_2} = \ket{0}$.
Since the initial state becomes a superposition of these eigenstates, $(\ket{0} + i\ket{1})/2$, when $\theta = \pi/2$, $Q_{ss}$ attains its maximum value at $a = 0$.
Thus,
\begin{equation}
\begin{split}
\mathrm{Tr}[\mathbf{Q}^{-1}] = \frac{1}{2t^2},
\end{split}
\label{TrQ_two_level}
\end{equation}
at the crossing point, i.e., $(s, a) = (0, 0)$.
As seen from Eq.~\eqref{Eq:Qss_linear} and Fig.~\ref{fig:QFI_2_z}(d) and (e), the crossing point serves as a hotspot for multi-parameter estimation, where both $Q_{ss}$ and $Q_{aa}$ attain their maximum values.
Since ${1}/{2t^2}$ is smaller than the value in Eq.~\eqref{TrQ_sigmaz}, and the hotspot remains unchanged with time, we choose $\theta = \pi/2$ for the sensing scheme discussed below.

%-----------------------figure 4-------------------------
\begin{figure*}[t]
\centering
\includegraphics[width=2.0\columnwidth]{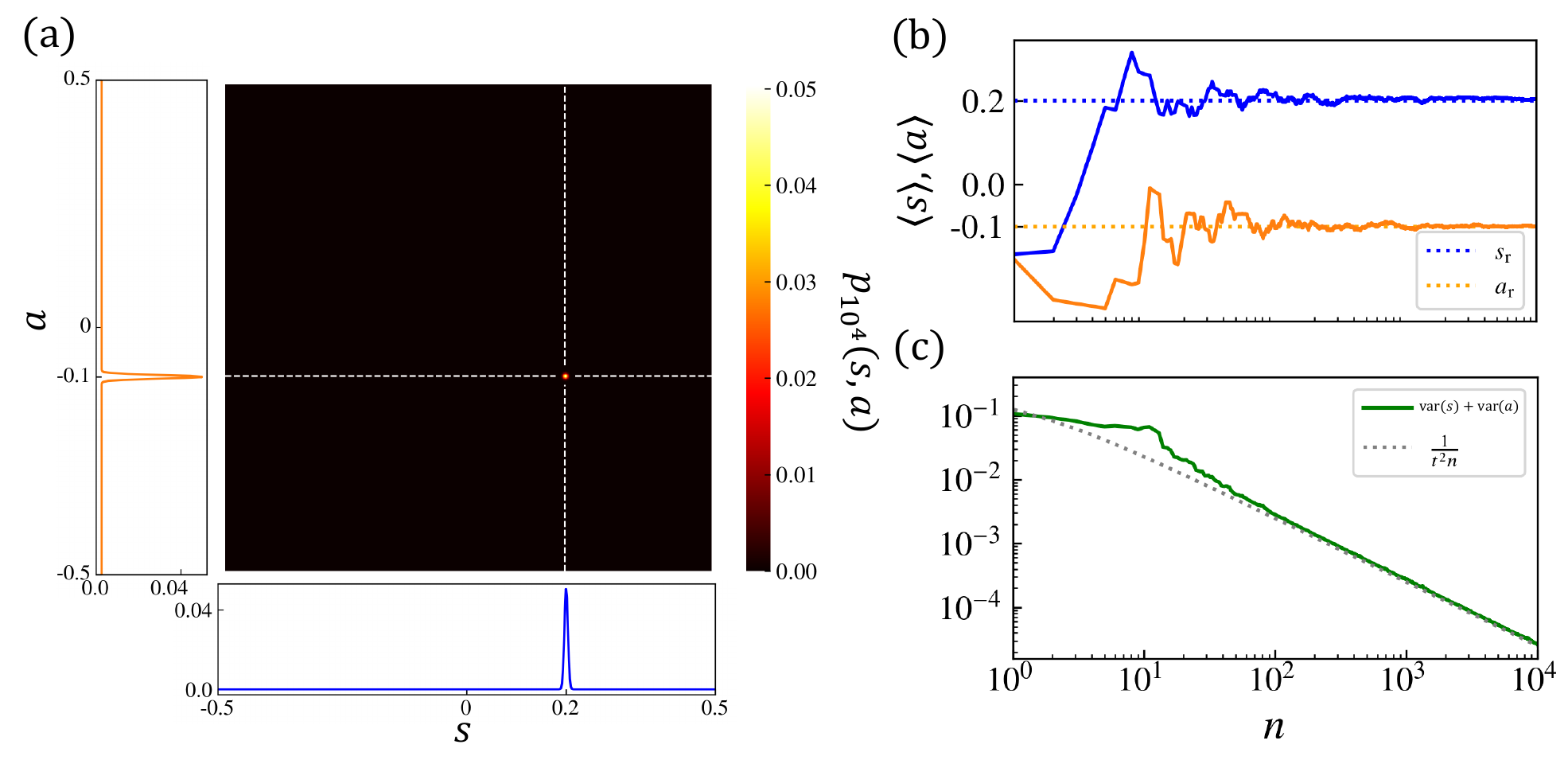}
\caption{(a)The probability density function $p_n(s,a)$ sharply peaked around the true value $(s_r,a_r) = (0.2, -0.1)$ where $n=10000$. 
(b)Blue and orange line exhibits the expected value of $s$ and $a$ as a function of the measurement number $n$. 
They approach to their true value represented the dotted line.
(c)Measured uncertainty of the parameters. 
Solid lines shows $\mathrm{var}(s)+\mathrm{var}(a)$ and the dot line shows the classical bound derived from the CFIM.}
\label{fig:Bayesian}
\end{figure*}
%-----------------------------------------------------

\subsection{Classical Fisher information}
In practical parameter estimation, one must specify a positive operator-valued measure (POVM) $\mathcal{M}$ corresponding to the measurement process.
For a given POVM, the lower bound on the variance of the estimated parameters is determined by the classical Fisher information matrix (CFIM) $\mathbf{C}$.
Each element of the CFIM is defined as
\begin{equation}
C_{\nu\mu} = \sum_m p(s,a|m) (\partial_\nu \ln p(s,a|m))(\partial_\mu \ln p(s,a|m))
\end{equation}
where $p(s,a|m) = \mathrm{Tr}[\ket{\psi} \bra{\psi} M_m]$ is the probability of obtaining the outcome $m$, with $\{\nu,\mu\} = \{s,a\}$.
The total estimation error is bounded as ${\rm var}(s) + {\rm var}(a) \ge \mathrm{Tr}[\mathbf{C}^{-1}]/n$, where $n$ is the number of measurements.

In multi-parameter estimation with non-commuting observables, no POVM can simultaneously attain the quantum limit derived by QFIM with pure state \cite{Matsumoto02,Ragy16}.
We therefore relax the requirement of achieving the quantum limit and consider the following convex combination of POVMs \cite{DAriano05}:
\begin{equation}
\mathcal{M} = p \mathcal{M}_a + (1-p) \mathcal{M}_s,
\end{equation}
where $\mathcal{M}_a = \{\ket{0}\bra{0}, \ket{1}\bra{1}\}$, $\mathcal{M}_s = \{\ket{+}\bra{+}, \ket{-}\bra{-}\}$, and $0\le p \le 1$.
The POVM $\mathcal{M}_a$ is optimal measurement operators for single-parameter estimation of $a$, while $\mathcal{M}_s$ is optimal for estimating $s$ at the crossing point (see Appendix~\ref{cal_CFI}).
$\mathcal{M}$ corresponds to a measurement process in which $\mathcal{M}_a$ is performed with classical probability $p$, while $\mathcal{M}_s$ is performed with probability $(1-p)$.
For $\theta = \pi/2$, the CFI at $s = s_c$, obtained from Eq.~\eqref{eq:Unitary_gene}, is given by
\begin{equation}
\begin{split}
    \mathrm{C}_{ss} &= 4 (1-p) t^2 {\rm sinc}^2(q t) \cos^2(q t) (\partial_s \delta)^2, \\
    \mathrm{C}_{aa} &= 4 p t^2 (\partial_a q)^2, \\
    \mathrm{C}_{sa} &= 0.
\end{split}
\end{equation}
Thus,
\begin{equation}
    \mathrm{Tr}[\bm{\mathrm{C}}^{-1}] = \frac{1}{4t^2}
    \frac{(1-p) \cos^2(q t) \mathrm{sinc}^2(q t) (\partial_s \delta)^2 + p (\partial_a q)^2}
    {p(1-p) \cos^2(q t) \mathrm{sinc}^2(q t) (\partial_s \delta)^2 (\partial_a q)^2}.
\end{equation}
If we choose
\begin{equation}\label{eq:POVM_prob}
p = \frac{|\partial_s \delta(s_c)|}{|\partial_s \delta(s_c)| + |\partial_a q(a_c)|},
\end{equation}
where $\partial_\mu f(\mu^*) = \frac{\partial f}{\partial \mu}\big|_{\mu=\mu^*}$,
the classical CR bound at the crossing point $(s,a) = (s_c,a_c)$ is given by
\begin{equation}\label{C_bound_two_level}
    \mathrm{Tr}[\bm{\mathrm{C}}^{-1}] = \frac{1}{4t^2} \left(\frac{1}{|\partial_s \delta(s_c)|} + \frac{1}{|\partial_a q(a_c)|}\right)^2.
\end{equation}
This value is equal to the Holevo CR bound \cite{Holevo11,Suzuki16}, indicating that the chosen POVM is optimal at the crossing point (see Appendix~\ref{Holevo}).
Using the Hamiltonian in Eq.~(\ref{simple_two}), we obtain
\begin{equation}
\mathrm{Tr}[\mathbf{C}^{-1}] = \frac{1}{t^2}.
\end{equation}

\subsection{Global sensing}
Next, we demonstrate global two-parameter estimation using the POVM $\mathcal{M}$ in the two-level system with $\theta = \pi/2$.
We consider two true parameters $s_r$ and $a_r$ located far from the estimation hot spot.
To improve estimation precision, we introduce control parameters, which shift the system toward the hot spot i.e., the crossing point.
However, adjusting control parameters is challenging in a global sensing setting without prior knowledge of the true parameters. 
To address this issue, we employ an adaptive strategy that combines feedback control with Bayesian estimation \cite{Salvia23,Yu25}, allowing the system to be steered toward the crossing point during the estimation process.

Later, we describe a measurement scheme based on Bayesian estimation.
In the following demonstration, the true parameter values to be estimated are $(s_r, a_r) = (0.2, -0.1)$, which are not close to the crossing point.
We assume prior knowledge of parameter ranges that contain the true values, i.e., $s_{\rm min} < s_r < s_{\rm max}$ and $a_{\rm min} < a_r < a_{\rm max}$.
Accordingly, we assume a uniform prior distribution, $p_0(s,a) = 1/{(s_{\rm max}-s_{\rm min})(a_{\rm max}-a_{\rm min})}$.
Our goal is to update the probability density function $p_n(s, a)$ at each step so that it becomes sharply peaked around the true values, starting from the prior distribution $p_0(s, a)$.
The $n$th probability density function $p_n(s, a)$ is updated from $p_{n-1}(s, a)$ according to the following rule.
First, the expectation values $\ev{s}_{n-1} = \int s p_{n-1}(s,a) ds da $ and $\ev{a}_{n-1} = \int a p_{n-1}(s,a) ds da $ are used to define the control parameters  $s_{\mathrm{ctrl}}^n = s_c - \ev{s}_{n-1}$ and $a_{\mathrm{ctrl}}^n = a_c - \ev{a}_{n-1}$ at each step, which shift the system toward the crossing point.
We then prepare the initial state $\ket{\psi_i}$ and evolve it under the Hamiltonian $H(s_r + s_{\mathrm{ctrl}}^n, a_r + a_{\mathrm{ctrl}}^n)$.
Then, we measure the evolved state $\ket{\psi_f^n}$ using the POVM $\mathcal{M}$ and obtain one of four possible outcomes, denoted by $m_n \in \{0,1,+,-\}$. 
We then update the probability density function using Bayes's theorem \cite{Sivia06}:
\begin{equation}
p_{n}(s,a) = \mathcal{N} p(m_n|s + s_{\mathrm{ctrl}}^n, a + a_{\mathrm{ctrl}}^n) p_{n-1}(s,a) ,
\end{equation}
where $p(m_n|s_r + s_{\mathrm{ctrl}}^n, a_r + a_{\mathrm{ctrl}}^n) = \mathrm{Tr}\big[\ket{\psi_f^n}\bra{\psi_f^n}M_{m_n}\big]$ represents the conditional probability of observing the outcome $m_n$ associated with the POVM element $M_{m_n}$, and $\mathcal{N} = \int p(s + s_{\mathrm{ctrl}}^n, a + a_{\mathrm{ctrl}}^n|m_n) p_{n-1}(s,a) ds da$ is the normalization constant.
The estimated parameters $\ev{s}_{n}$ and $\ev{a}_{n}$ are then updated accordingly.

The numerical simulation of the Bayesian adaptive strategy is shown in Fig.~\ref{fig:Bayesian}.
Figure~\ref{fig:Bayesian}(a) shows the probability density function $p_{10^4}(s,a)$ after $10^4$ update steps.
This distribution exhibits a sharp peak at the true parameter values. 
Although the initial distribution $p_{0}(s,a)$ was uniform, it became progressively narrower and more sharply peaked as the number of update steps increased.
Indeed, as the number of measurements increases, $\ev{s}_{n}$ and $\ev{a}_{n}$ converges to the true values $(s_r, a_r) = (0.2, -0.1)$ as shown in Fig.~\ref{fig:Bayesian}(b).
Meanwhile, as the effective parameters applied to the system, $(s_r + s_{\mathrm{ctrl}}^n, a_r + a_{\mathrm{ctrl}}^n)$, approach the crossing point, the measurement sensitivity is progressively enhanced.
This trend is further supported by the rapid decrease in the uncertainty of $p_{n}(s,a)$.
The estimation precision can be evaluated by the variance of the posterior distribution:
\begin{equation}
    \mathrm{var}(s) + \mathrm{var}(a) = \int p_{n}(s,a)\{(s-\ev{s})^2 + (a - \ev{a})^2\}ds da.
\end{equation}
Figure~\ref{fig:Bayesian}(c) shows the sum of the estimation variances, $\mathrm{var}(s)+\mathrm{var}(a)$, together with the classical Cramér--Rao bound $1/t^2n$ obtained from the CFIM at the crossing point. As the number of adaptive measurements increases, $\mathrm{var}(s)+\mathrm{var}(a)$ decreases and gradually approaches the classical bound.
%-----------------------figure 5-------------------------
\begin{figure}[t]
\centering
\includegraphics[width=0.8\columnwidth]{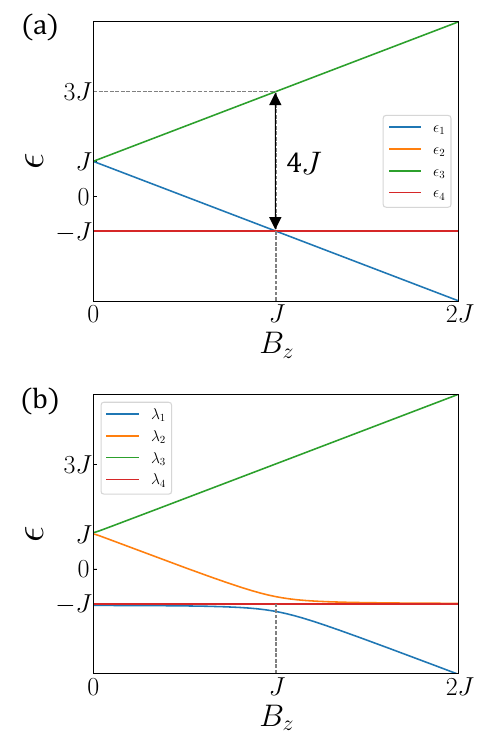}
\caption{The eigenvalues of Hamiltonian for (a)$B_x = 0$ and (b)$B_x = 0.2$. The energy levels of $\epsilon_2$ and $\epsilon_4$ are degenerate when $B_x=0$.}
\label{fig:Eigenvalue_Ising}
\end{figure}
%--------------------------------------------------------

%-----------------------figure 6-------------------------
\begin{figure*}[t]
\centering
\includegraphics[width=2.1\columnwidth]{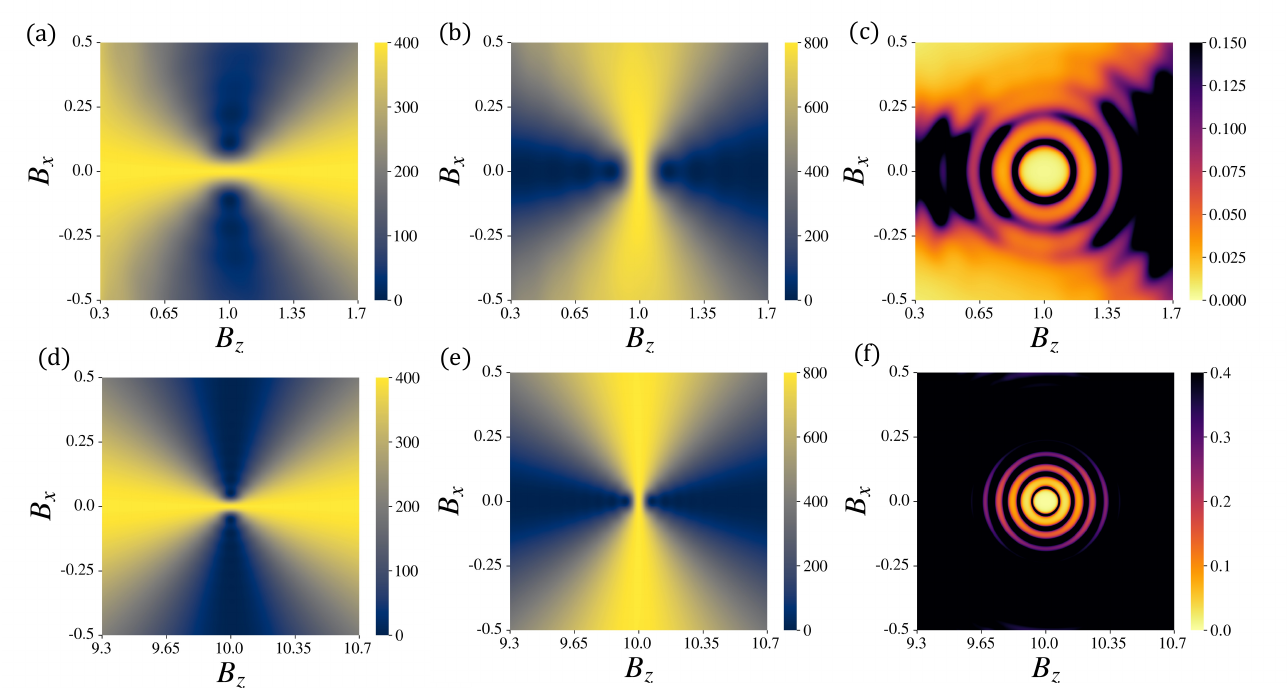}
\caption{The heatmaps of (a) $Q_{B_zB_z}$, (b) $Q_{B_xB_x}$, and (c) $\mathrm{Tr}[\mathbf{Q}^{-1}]$ for $\theta=\varphi=\pi/2$ and $J=1$.
The heatmaps of (d) $Q_{B_zB_z}$, (e) $Q_{B_xB_x}$, and (f) $\mathrm{Tr}[\mathbf{Q}^{-1}]$  for $\theta=\varphi=\pi/2$ and $J=10$.}
\label{fig:QFI_unitary_Ising}
\end{figure*}
%---------------------------------------------------------
%-----------------------figure 7-------------------------
\begin{figure*}[t]
\centering
\includegraphics[width=2.0\columnwidth]{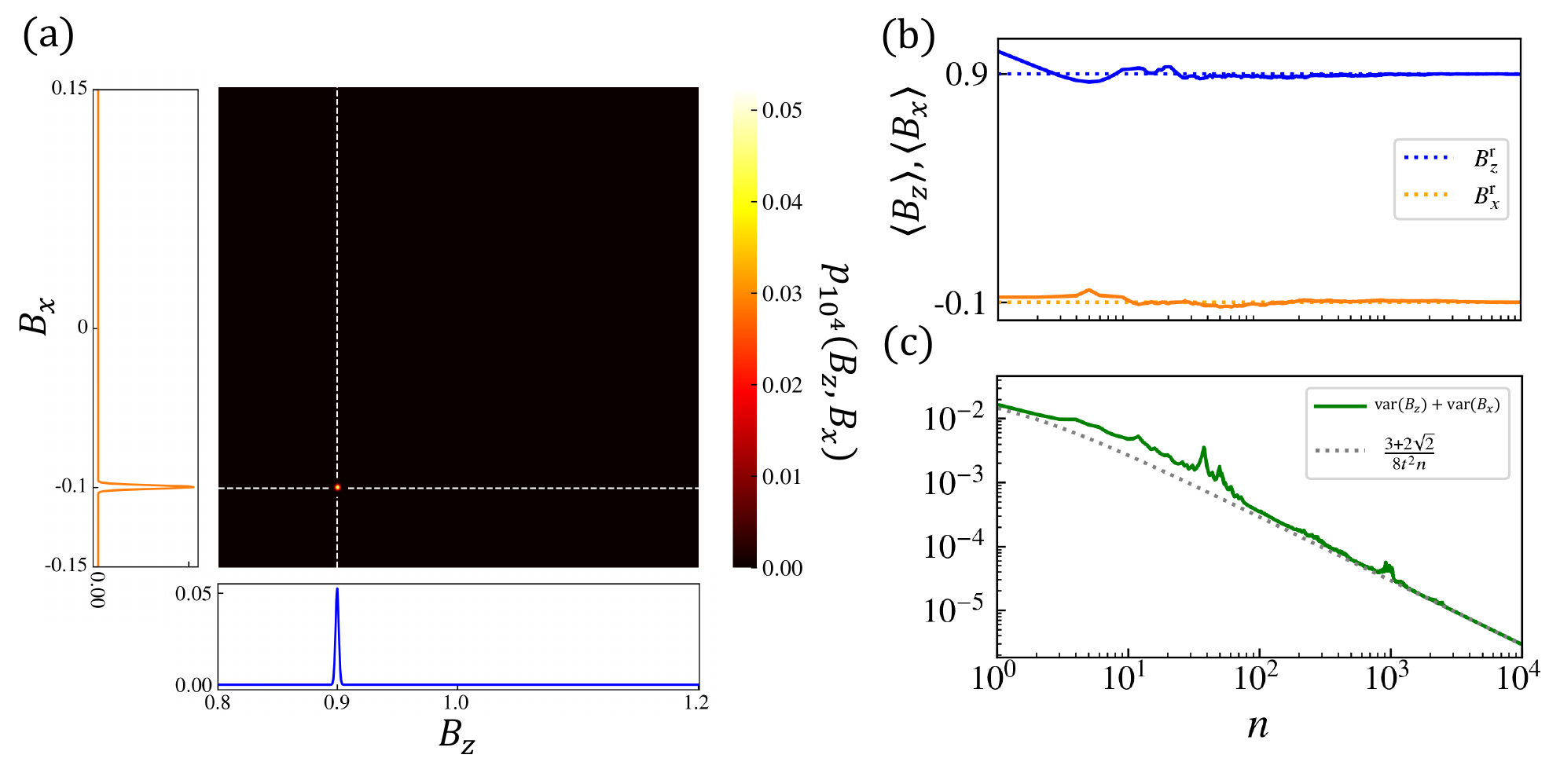}
\caption{(a)The probability density function $p_n(B_z,B_x)$ sharply peaked around the true value $(B_z^r,B_x^r) = (0.9, -0.1)$ where $n=10000$. 
(b)Blue and orange line exhibits the expected value of $B_z$ and $B_x$ as a function of the measurement number $n$. 
They approach to their true value represented the dotted line.
(c)Measured uncertainty of the parameters. 
Solid lines shows the variance of the vector $(Bz,Bx)$ and the dot line shows the classical bound derived from the CFIM.}
\label{fig:Bayesian_Ising}
\end{figure*}
%-----------------------------------------------------

\section{Ising model}\label{Sec:Ising}
\subsection{QFI with unitary process}
In this section, we examine the conditions under which the two-level approximation and parameter estimation perform well, using the two-spin Ising model.
In this system, the Hamiltonian is given by
\begin{equation}
\begin{split}
    H &= J\hat{\sigma}_{1z} \cdot \hat{\sigma}_{2z} + B_z (\hat{\sigma}_{1z} + \hat{\sigma}_{2z}) + B_x (\hat{\sigma}_{1x} + \hat{\sigma}_{2x})\\ &=
\begin{pmatrix}
J + 2B_z & B_x & B_x & 0 \\
B_x & -J & 0 & B_x \\
B_x & 0 & -J & B_x \\
0 & B_x & B_x & J - 2B_z 
\end{pmatrix}.
\end{split}
\label{Eq:Ising_Hamiltonian}
\end{equation}
The parameters to be estimated are $B_z$ and $B_x$.
The Hamiltonian has four eigenvalues $\lambda_i$ and corresponding eigenvectors $\ket{\lambda_i}$.
Figure~\ref{fig:Eigenvalue_Ising} shows the eigenvalues of the Hamiltonian for (a)$B_x=0$ and (b)$B_x=0.2$.
For $B_x=0$, a level crossing occurs at $B_z=J$, whereas for finite $B_x$, the levels $\lambda_1$ and $\lambda_2$ exhibit level repulsion.
The parameter $B_x$ induces level repulsion and plays a role similar to that of the parameter $a$ in the two-level system discussed above.
To rewrite the Hamiltonian, we first consider its eigenvalues at $B_x = 0$, which are given by $\epsilon_1 = J-2B_z, \epsilon_2 = -J , \epsilon_3 = J + 2B_z$, and $\epsilon_4 = -J$.
The corresponding eigenvectors are
\begin{equation}
\begin{split}
    \ket{v_1} =
\begin{pmatrix}
0 \\
0\\
0\\
1 
\end{pmatrix}
    \ket{v_2} =\frac{1}{\sqrt{2}}
\begin{pmatrix}
0\\
1\\
1\\
0
\end{pmatrix}
    \ket{v_3} =
\begin{pmatrix}
1 \\
0\\
0\\
0
\end{pmatrix}
    \ket{v_4} =\frac{1}{\sqrt{2}}
\begin{pmatrix}
0\\
1\\
-1\\
0
\end{pmatrix}
\end{split}
\label{eigen_vec}
\end{equation}
Then, we rewrite the Hamiltonian in Eq.~\eqref{Eq:Ising_Hamiltonian} using the basis $\{ \ket{v_1}, \ket{v_2}, \ket{v_3}, \ket{v_4} \}$
\begin{equation}
    H=\begin{pmatrix}
J - 2B_z & \sqrt{2}B_x & 0 & 0 \\
\sqrt{2}B_x & -J & \sqrt{2}B_x & 0 \\
0 & \sqrt{2}B_x & J + 2B_z & 0 \\
0 & 0 & 0 & -J
\end{pmatrix}
= H_{3 \times3} \oplus (-J) .
\label{Ising_Hamiltonian_2}
\end{equation}
The Hamiltonian can be decomposed into a direct sum of subspaces spanned by $\{{\ket{v_1}, \ket{v_2}, \ket{v_3}}\}$ and $\{{\ket{v_4}}\}$ for arbitrary values of $B_x$, $B_z$, and $J$.
Therefore, as long as the initial state lies in the subspace spanned by $\{{\ket{v_1}, \ket{v_2}, \ket{v_3}}\}$, the dynamics under unitary evolution can be described by $H_{3 \times3}$.

Next, we approximate $H_{3 \times3}$ by a two-dimensional Hamiltonian around the level-repulsion point at $B_z = J$.
The effective Hamiltonian in the subspace spanned by $\ket{v_1}$ and $\ket{v_2}$ is 
\begin{equation}
\begin{split}
    H_{\mathrm{eff}} & = \begin{pmatrix}
J - 2B_z & \sqrt{2} B_x\\
 \sqrt{2} B_x & -J\\
\end{pmatrix}\\
& = - B_z{\sigma}_I + (J - B_z) {\sigma}_z + \sqrt{2} B_x {\sigma}_x.
\end{split}
\label{eff_Hamiltonian}
\end{equation}
Except for the constant energy shift $-B_z\sigma_I$, this Hamiltonian is obtained by setting $s= B_z$, $a=B_x$, $\delta = (J-B_z)$, and $q = \sqrt{2}B_x$ in Eq.~(\ref{Eq:general_Hamiltonian}).
As the two-level system, we set the initial state as the superposition of the two basis of the effective Hamiltonian $\ket{v_1}$ and $\ket{v_2}$ as
\begin{equation}
    \ket{\psi_i} = \cos{\left( \frac{\theta}{2}\right)}\ket{v_1} + e^{i\varphi}\sin{\left( \frac{\theta}{2}\right)} \ket{v_2}.
\end{equation}
As same as Eq.~(\ref{Eq:QFI_two_2pi}), in the level-repulsion regime around $B_z = J$, the QFI are obtained as
\begin{equation}
\begin{split}
    \mathrm{Q}_{B_z B_z} &=  4 t^2\mathrm{sinc}^2(\sqrt{2} B_x t)\{1 - g^2\}, \\
    \mathrm{Q}_{B_x B_x} &=  8t^2(1-\ev{\sigma_x'}), \\
    \mathrm{Q}_{B_z B_x} &= \mathrm{Q}_{B_x B_z} = - 4 \sqrt{2} t^2 g \mathrm{sinc}(\sqrt{2} B_x t) \ev{\sigma_x'}. 
\end{split}
\end{equation}
where $g = \sin(\sqrt{2} B_x t) \ev{\sigma'_y} + \cos(\sqrt{2} B_x t) \ev{\sigma'_z}$ with $\sigma'_x = \ket{v_1}\bra{v_2} + \ket{v_2}\bra{v_1}$, $\sigma_y' = -i\ket{v_1}\bra{v_2} + \ket{v_2}\bra{v_1}$, and $\sigma'_z = \ket{v_1}\bra{v_1} - \ket{v_2}\bra{v_2}$.
At the crossing point $(B_z, B_x) = (J, 0)$ and $\theta=\varphi=\pi/2$, the QFI are 
\begin{equation}
\begin{split}
    \mathrm{Q}_{B_z B_z} &=  4 t^2, \\
    \mathrm{Q}_{B_x B_x} &=  8t^2, \\
    \mathrm{Q}_{B_z B_x} &= 0. 
\end{split}
\end{equation}
Then, we have $\mathrm{Tr[\mathbf{Q}^{-1}]} = 3/8t^2$.

Figures~\ref{fig:QFI_unitary_Ising}(a)–(c) show the numerical results for $Q_{ss}$, $Q_{aa}$, and $\mathrm{Tr}[\bm{\mathrm{Q}}^{-1}]$ under unitary evolution at time $t = 10$ with $\theta=\varphi=\pi/2$ and $J=1$.
In particular, near the center of the heat map of $\mathrm{Tr}[\mathbf{Q}^{-1}]$, a concentric-ring structure similar to that observed in Fig.~\ref{fig:QFI_2_z}(f) can be seen. 
However, this structure becomes increasingly blurred toward the upper and lower regions of the figure.
Figures~\ref{fig:QFI_unitary_Ising}(d)–(f) show the numerical results for $Q_{ss}$, $Q_{aa}$, and $\mathrm{Tr}[\bm{\mathrm{Q}}^{-1}]$ with $\theta=\varphi=\pi/2$ and $J=10$.
For both $J = 1$ and $J = 10$, $Q_{B_x B_x}$ and $Q_{B_z B_z}$ take their maximum values at the level-crossing point.
These maximum values match the theoretical predictions $\mathrm{Q}_{B_z B_z} = 4t^2 =400$ and $\mathrm{Q}_{B_x B_x} = 8t^2= 800$.
The two-level approximation works well when the subspace spanned by the two eigenstates $\ket{\lambda_1}$ and $\ket{\lambda_2}$ is close to the subspace spanned by $\ket{v_1}$ and $\ket{v_2}$.
In particular, for the case of $J = 10$, the two-level approximation is valid (see App.~\ref{subspace}), and the QFI behaves similarly to that of the two-level system shown in Fig.~\ref{fig:QFI_2_z}(d)–(f).
In the Ising model, increasing the exchange interaction enlarges the energy gap between the crossing point and the other levels.
However, the case of $J = 1$, shown in Fig.~\ref{fig:QFI_unitary_Ising}(c), provides a wider parameter region with low $\mathrm{Tr}[\mathbf{Q}^{-1}]$ and is therefore more suitable for practical parameter estimation.
Next, we demonstrate the global sensing scheme in the Ising model for $J=1$.

\subsection{CFIM and global sensing scheme}
This section demonstrates that the sensing method developed for a two-level system can be extended to systems exhibiting level crossings, such as the Ising model.
To this end, it suffices to replace $\ket{0}$ with $\ket{v_1}$ and $\ket{1}$ with $\ket{v_2}$ in the method described in Appendix~\ref{cal_CFI}.
Therefore, the initial state is $\ket{v_1} + i\ket{v_2}$ i.e. $\theta=\varphi=\pi/2$.
The POVM is given by $\mathcal{M} = \{\ket{u_m} \bra{u_m} \}^8_{m=1}$ with $\ket{u_1} = \sqrt{p} \ket{v_1}$, $\ket{u_2} = \sqrt{p} \ket{v_2}$, $\ket{u_3} = \sqrt{1-p} (\ket{v_1} + \ket{v_2})/2$, $\ket{u_4} = \sqrt{1-p} (\ket{v_1} - \ket{v_2})/2$, $\ket{u_5} = \sqrt{p} \ket{v_3}$, $\ket{u_6} = \sqrt{p} \ket{v_4}$, $\ket{u_7} = \sqrt{1-p}(\ket{v_3} + \ket{v_4})/2$, and $\ket{u_8} = \sqrt{1-p}(\ket{v_3} - \ket{v_4})/2$.
The first four correspond to the POVM elements chosen with reference to the two-level system, whereas the remaining four are introduced to satisfy the completeness condition of the POVM.
Using Eq.~\eqref{eq:POVM_prob}, we choose $p=1/(1+\sqrt{2})$.
Figure~\ref{fig:Bayesian_Ising} shows the numerical result of the Bayesian adaptive strategy for estimating two magnetic fields, $B_z$ and $B_x$.
The probability density function $p_{10^4}(s,a)$ after $10^4$ update steps is shown in Fig.~\ref{fig:Bayesian_Ising}(a).
This distribution exhibits a sharp peak at the true values of the magnetic fields. 
Indeed, as the number of measurements increases, $\ev{B_z}$ and $\ev{B_x}$ converge to the true value $(B_z^r, B_x^r) = (0.9, -0.1)$ as shown in Fig.~\ref{fig:Bayesian_Ising}(b).
Figure~\ref{fig:Bayesian_Ising}(c) shows that the estimation precision $\mathrm{var}(B_z) + \mathrm{var}(B_x)$ decreases as the number of adaptive measurements increases.
In this case, $\mathrm{var}(B_z) + \mathrm{var}(B_x)$ approaches the classical bound derived in Eq.~\eqref{C_bound_two_level}, $$\frac{3+2\sqrt{2}}{8t^2n},$$ at the crossing point.
By focusing on energy-level intersections in this way, we can construct a unified parameter-estimation scheme that is applicable regardless of the overall structure of the Hamiltonian.

\section{Conclusion}\label{Sec:Conclusion}
In conclusion, we have analyzed systems exhibiting energy-level crossings to investigate how estimation precision behaves in multi-parameter quantum estimation. 
For a minimal two-parameter model described by a two-dimensional real symmetric Hamiltonian, the QFI for both parameters simultaneously diverge at the level-crossing point when the ground state is used as the probe state.
However, multi-parameter estimation is impossible because the QFIM is a singular matrix.
Thus, we focused on unitary time evolution for multi-parameter estimation.
Adjusting the initial state, $\mathrm{Tr[\mathbf{Q}^{-1}]}$ attains their minima at the crossing point.
We constructed a sensing scheme that attains the Holevo Cramér–Rao bound by combining the optimal POVM with a Bayesian adaptive method.
To investigate the conditions under which this measurement scheme performs well and how to construct the measurement scheme, we considered another system exhibiting level crossings, namely the two-spin Ising model.
The behavior of the QFI around $B_z=J$ approaches that of the two-level model as the hybridization with the additional energy level is reduced. 
Interestingly, however, the influence of the additional level can also have a beneficial effect on multi-parameter estimation, as manifested by the reduction in $\mathrm{Tr}[\mathbf{Q}^{-1}]$ away from the level-crossing point. 
Taken together, these results demonstrate that the present work is not merely a study of parameter estimation in a particular qubit model, but reveals how level crossings and coupling to additional energy levels jointly determine the performance of multi-parameter quantum estimation. 
In particular, the local level-crossing structure of a Hamiltonian provides natural ``hot spots'' for high-sensitivity quantum metrology. 
Because level crossing structures are ubiquitous across a wide range of quantum systems, these findings provide a general framework for the development of simple and scalable quantum sensors based on level crossings.

\section*{Acknowledgments}
This work was supported by JSPS KAKENHI Grant Numbers 26KJ0593, 24K00563, 24K02948, and 24K06878. 
R. T. and H. M. thank to financial support from GP-Spin program at Tohoku University. 
L.B.H. acknowledges support from the Tohoku Initiative for Fostering Global Researchers for Interdisciplinary Sciences (TI-FRIS) of MEXT’s Strategic Professional Development Program for Young Researchers.
H. M. acknowledges support from CSIS at Tohoku University.

\appendix

\section{QFI at ground state}\label{cal_QFI_ground}
For calculate the QFIM, we differentiate the ground state $\ket{\eta_1}$ by parameters $s$ and $a$
\begin{equation}
    \ket{\partial_s \eta_1} = \frac{1}{2}\partial_s{\phi}\left(\cos \frac{\phi}{2} \ket{0} + \sin \frac{\phi}{2} \ket{1} \right),
\end{equation}
\begin{equation}
    \ket{\partial_a \eta_1} = \frac{1}{2}\partial_a{\phi} \left(\cos \frac{\phi}{2} \ket{0} + \sin \frac{\phi}{2} \ket{1} \right).
\end{equation}
Then,
\begin{equation}
    \braket{\partial_s \eta_1} = \frac{1}{4}(\partial_s{\phi})^2, \    \braket{\partial_a \eta_1} = \frac{1}{4}(\partial_a{\phi})^2, 
\end{equation}
\begin{equation}
    \braket{\partial_{s (a)} \eta_1}{\eta_1}
 = \frac{\partial_{s (a)}{\phi}}{2} \left(\frac{1}{2} (\sin{\phi} - \sin{\phi})\right) = 0. 
\end{equation}

The elements in QFIM 
\[
Q_{\mu\nu} = 
4\mathrm{Re}\Big(\braket{\partial_\mu \eta_1}{\partial_\nu \eta_1} - \braket{\partial_\mu \eta_1}{\eta_1}\braket{\eta_1}{\partial_\nu  \eta_1}\Big)
\]
explicitly give 
\begin{equation}
    \mathrm{Q}_{ss} = (\partial_s{\phi})^2 =\frac{q^2}{(\delta^2 + q^2)^2} (\partial_s \delta)^2,
\end{equation}
\begin{equation}
    \mathrm{Q}_{aa} = (\partial_a{\phi})^2 =\frac{\delta ^2}{(\delta^2 + q^2)^2}(\partial_a q)^2, 
\end{equation}
\begin{equation}
    \mathrm{Q}_{sa} = \mathrm{Q}_{as} = (\partial_s{\phi})(\partial_a{\phi}) =-\frac{\delta q}{(\delta^2 + q^2)^2} \partial_s \delta \partial_a q.
\end{equation}
For fixed \(q\) and finite \(\partial_s\delta\), \(Q_{ss}\) is maximized at \(\delta(s_c)=0\), corresponding to the minimum energy gap. At this point, \(
Q_{ss}(s_c)=\frac{(\partial_s\delta)^2}{q^2},
\)
which diverges as \(q\rightarrow0\). Similarly, for fixed nonzero \(\delta\), \(Q_{aa}\) is maximized at \(q(a)=0\), where \(
Q_{aa}=\frac{(\partial_aq)^2}{\delta^2},
\)
and diverges as \(\delta\rightarrow0\). The point \(\delta=q=0\) corresponds to an exact level crossing, where the ground state becomes degenerate and the above nondegenerate-ground-state QFI expressions are not well defined.

%For fixed $q$ and a finite $\partial_s \delta$, $Q_{ss}$ is maximized at $\delta(s) = 0$, i.e., $s=s_c$. 
%%If $\partial_s \delta \neq 0$ around $\delta(s) = 0$, $Q_{ss}$ achieves maximum value at $\delta(s) = 0$, i.e., $s=s_c$. 
%Keeping $s=s_c$ while reducing $q$ to 0, the $Q_{ss}$ diverges as mention above. And $Q_{aa}$ achieves maximum value at $q(a) = 0$, corresponding to the vanishing level-repulsion coupling. Both $Q_{ss}$ and $Q_{aa}$ diverge at $\delta = q = 0$.

\section{Quantum metrology at the evolved state}\label{App:QFI_Unitary}
Let us start from the unitary operator $U=e^{-iHt}$ expressed by the Pauli matrix
\begin{equation}
    U = e^{-iHt} = r_I \sigma_I + i\bm{r} \cdot\bm{\sigma}
\end{equation}
where $r_I = \cos(\Omega t)$ with $\Omega = \sqrt{\delta^2 + q^2}$, $\bm{r} = (r_x, r_y, r_z)$ as 
\begin{equation}
    r_x = - \frac{q}{\Omega} \sin(\Omega t); \
    r_y = 0; \
    r_z = - \frac{\delta}{\Omega} \sin(\Omega t).
\end{equation}
Under this transformation, an initial state $\ket{\psi_i}$ evolves to a final state $\ket{\psi_f} = U\ket{\psi_i}$.

\begin{itemize}
    \item To obtain $\mathbf{Q}$ at the final state, we first derive
\end{itemize}
\begin{equation}
    \partial_s U = \frac{\partial r_I}{\partial s} \sigma_I + i\Big(\frac{\partial \bm{r}}{\partial s} \cdot \bm{\sigma}\Big) = \mu_I \sigma_I + i\bm{\mu} \cdot \bm{\sigma},
\end{equation}
where 
\begin{equation}\label{final_state_s}
    \begin{split}
        \mu_I &\equiv \frac{\partial r_I}{\partial s} = -\frac{t\delta}{\Omega} \sin(\Omega t)\ \partial_s{\delta},\\
        \mu_x &\equiv \frac{\partial r_x}{\partial s} = \frac{q \delta}{\Omega^3} \big[\sin(\Omega t) - \Omega t\cos(\Omega t)\big]\ \partial_s{\delta}, \\
        \mu_y &\equiv \frac{\partial r_y}{\partial s} = 0,\\
        \mu_z &\equiv \frac{\partial r_z}{\partial s} = - \frac{1}{\Omega^3} \big[q^2 \sin(\Omega t) + \Omega t\delta^2  \cos(\Omega t)\big]\ \partial_s{\delta}.
    \end{split}
\end{equation}

\begin{itemize}
    \item Then, we also derive
\end{itemize}
\begin{equation}
    \partial_a U = \frac{\partial r_I}{\partial a} \sigma_I + i\Big(\frac{\partial \bm{r}}{\partial a} \cdot \bm{\sigma}\Big) = \nu_I \sigma_I + i\bm{\nu} \cdot \bm{\sigma},
\end{equation}
where 
\begin{equation}\label{final_state_a}
    \begin{split}
        \nu_I &= - \frac{qt}{\Omega} \sin(\Omega t)\partial_a{q},\\
        \nu_x &= - \frac{1}{\Omega^3} \big[\delta^2 \sin(\Omega t) + \Omega tq^2  \cos(\Omega t)\big]\partial_a{q}, \\
        \nu_y &= 0,\\
        \nu_z &= \frac{q\delta}{\Omega^3} \big[\sin(\Omega t) - \Omega t\cos(\Omega t)\big]\partial_a{q}.
    \end{split}
\end{equation}

\begin{itemize}
    \item Next, we calculate the first part of the QFIM as 
\end{itemize}
\begin{equation}
    \bra{\psi_i} \partial_s U^{\dagger}\partial_s U \ket{\psi_i} =\mu_I^2 + \mu_x^2 + \mu_z^2,
\end{equation}
and
\begin{equation}
    \bra{\psi_i} \partial_a U^{\dagger}\partial_a U \ket{\psi_i} =\nu_I^2 + \nu_x^2 + \nu_z^2,
\end{equation}
and
\begin{equation}
\begin{split}
    \mathrm{Re}[\bra{\psi_i} \partial_s U^{\dagger}\partial_a U \ket{\psi_i}] &= \mathrm{Re}[\bra{\psi_i} \partial_a U^{\dagger}\partial_s U \ket{\psi_i}] \\
    & = \mu_I\nu_I+ \mu_x\nu_x + \mu_z\nu_z.
\end{split}
\end{equation}

\begin{itemize}
    \item The second part of the QFIM gives
\end{itemize}
\begin{equation}
\begin{split}
        |\braket{\partial_s \psi_f}{\psi_f}|^2 = j_I^2 + 
        \sum_{i,k} j_ij_k \ev{\sigma_i}\ev{\sigma_k},
        %\big(j_x \ev{\sigma_x} + j_y\ev{\sigma_y}+ j_z\ev{\sigma_z}\big)^2,
\end{split}
\end{equation}
where $\ev{\sigma_k} = \bra{\psi_i} \sigma_k\ket{\psi_i}, j_I= r_I\mu_I + r_x\mu_x+ r_z\mu_z$, and
\begin{equation}
    \begin{split}
        j_x = r_I\mu_x - r_x \mu_I,\\
        j_y = r_z\mu_x - r_x \mu_z,\\
        j_z = r_I\mu_z - r_z \mu_I,
    \end{split}
\end{equation}
and
\begin{equation}
\begin{split}
        |\braket{\partial_a \psi_f}{\psi_f}|^2 = l_I^2 + \sum_{i,k} l_il_k \ev{\sigma_i}\ev{\sigma_k},
        %\big(l_x \ev{\sigma_x} + l_y\ev{\sigma_y} + l_z\ev{\sigma_z}\big)^2,
\end{split}
\end{equation}
where $l_I= r_I\nu_I + r_x\nu_x+ r_z\nu_z$, and
\begin{equation}
    \begin{split}
        l_x = r_I\nu_x - r_x \nu_I,\\
        l_y = r_z\nu_x - r_x \nu_z,\\
        l_z = r_I\nu_z - r_z \nu_I.
    \end{split}
\end{equation}
and 
\begin{equation}
        \mathrm{Re}[\braket{\partial_s \psi_f}{\psi_f}\braket{\psi_f}{\partial_a \psi_f}] 
        %&= \mathrm{Re}[\bra{\psi_i} \partial_s U^\dagger U \ket{\psi_i} \bra{\psi_i} U^\dagger  \partial_aU \ket{\psi_i}] \\
        = j_I l_I + \sum_{i,k} j_il_k\ev{\sigma_i}\ev{\sigma_k}.
        %\big(j_x \ev{\sigma_x} + j_y \ev{\sigma_y} + j_z \ev{\sigma_z}\big)\\
        %& \qquad \cdot \big(l_x \ev{\sigma_x} + l_y \ev{\sigma_y} + l_z \ev{\sigma_z}\big)
\end{equation}

\begin{itemize}
    \item $s\approx s_c$, we have $\delta \rightarrow 0$. Let $f = \sin(qt) \ev{\sigma_y} + \cos(qt) \ev{\sigma_z}$, and define $\mathrm{sinc}(qt) = \frac{\sin(qt)}{qt}$, the QFI becomes
\end{itemize}
\begin{align}
\notag \mathrm{Q}_{ss} &= 4 t^2 {\rm sinc}^2(qt)\Big[1 - f^2\Big] (\partial_s\delta)^2,\\
\notag \mathrm{Q}_{aa} &= 4t^2\Big(1 - \ev{\sigma_x}^2\Big)(\partial_a{q})^2,\\
\notag \mathrm{Q}_{sa} &=\mathrm{Q}_{as} = -4t^2f {\rm sinc}(qt)\ev{\sigma_x} \partial_s\delta\partial_a{q}.
\end{align}
\begin{itemize}
    \item Finally, we also evaluate $\mathrm{Tr}[\mathbf{Q}^{-1}]$ as
\end{itemize}
\begin{equation}
    \mathrm{Tr}[\mathbf{Q}^{-1}] = \frac{\mathrm{sinc}^2(q t) \big(1 - f^2\big)(\partial_s\delta)^2 + (1-\ev{\sigma_x}^2) ((\partial_aq)^2)}{4t^2 \mathrm{sinc}^2(q t) \big(1-f^2-\ev{\sigma_x}^2\big)(\partial_s\delta)^2( \partial_aq)^2}
\label{Q-1-compact}
\end{equation}

\section{Holevo-Cramér-Rao bound}\label{Holevo}
In multi-parameter quantum estimation, the ultimate precision limit is characterized by the Holevo bound \cite{Holevo11,Suzuki16}.
Unlike the quantum Cramér–Rao bounds based on the SLD, which may be incompatible for noncommuting observables, the Holevo bound provides the tightest achievable bound on the mean-square error.
To calculate the Holevo bound in our model, we define a set of $2 \times 2$ Hermitian operators $\mathcal{X}$ satisfying
\begin{equation}
\begin{split}
    \mathcal{X} \coloneqq \{\vec{X} =(X^{s},X^{a})| &\forall i \ \mathrm{Tr}[\rho X^i] = 0, \\
    & \forall i,j \ \mathrm{Tr}[\frac{\partial \rho}{\partial i}X^j] = \delta_{ij}\}, 
\end{split}
\end{equation}
where $i,j = s,a$ and $\rho$ is the quantum state.
The Holevo function is defined by 
\begin{equation}
    h(\vec{X}) \coloneqq \mathrm{Tr}[\operatorname{Re} Z(\vec{X})] + \| \operatorname{Im} Z(\vec{X}) \|_1,
\end{equation}
where the $2 \times 2$ Hermitian matrix $Z(\vec{X})$ is given
\begin{equation}
    Z(\vec{X})_{ij} \coloneqq \mathrm{Tr}[\rho X^i X^j]
\end{equation}
and $\|A\|_1 \coloneqq \mathrm{Tr}[\sqrt{A^{\dagger}A}]$ denotes the trace norm.
The Holevo bound is defined through the following optimization
\begin{equation}
    C^H \coloneqq \min_{\vec{X} \in \mathcal{X}} h(\vec{X}).
\end{equation}

We have $\rho = \ket{\psi_i} \bra{\psi_i}$, since we only consider the case at the crossing point.
Therefore, from Eq.~\eqref{final_state_s} and Eq.~\eqref{final_state_a} we have $\frac{\partial \rho}{\partial s} =  -\partial_s\delta(s_c)t\sigma_x$ and $\frac{\partial \rho}{\partial a} = \partial_a q(a_c)t\sigma_z$ where $\partial_i f(i^*) = \frac{d f}{d i}|_{i=i^*}$.
The explicit form of $(X^s,X^a)$ becomes
\begin{equation}
\begin{split}
    &X^s = \begin{pmatrix}
        u & -\frac{1}{2t\partial_s\delta(s_c)}+ ui \\
        -\frac{1}{2t\partial_s\delta(s_c)}- ui & u
    \end{pmatrix} \\
    &X^a = \begin{pmatrix}
        v & (v - \frac{1}{2t\partial_a q(a_c)})i \\
        -(v - \frac{1}{2t\partial_a q(a_c)})i & v - \frac{1}{t\partial_a q(a_c)}
    \end{pmatrix}
\end{split}
\end{equation}
where $u,v \in \mathbb{R}$.
Then, 
\begin{equation}
    Z(\vec{X}) = \begin{pmatrix}
        \frac{1}{4t^2\{\partial_s\delta(s_c)\}^2} & \frac{i}{4t^2\partial_s\delta(s_c)\partial_aq(a_c)}\\
        -\frac{i}{4t^2\partial_s\delta(s_c)\partial_aq(a_c)} & \frac{1}{4t^2\{\partial_aq(a_c)\}^2}
    \end{pmatrix}.
\end{equation}
We have
\begin{equation}
\begin{split}
    h(\vec{X}) & = \mathrm{Tr} \left[\begin{matrix}
        \frac{1}{4t^2\{\partial_s\delta(s_c)\}^2} & 0\\
        0 & \frac{1}{4t^2\{\partial_aq(a_c)\}^2}
    \end{matrix}\right] \\
    & \qquad \ \ \ + \left\|\begin{matrix}
        0 & \frac{1}{4t^2\partial_s\delta(s_c)\partial_aq(a_c)} \\
        -\frac{1}{4t^2\partial_s\delta(s_c)\partial_aq(a_c)} & 0
    \end{matrix}\right\|\\
    &= \frac{1}{4t^2\{\partial_s\delta(s_c)\}^2} + \frac{1}{4t^2\{\partial_aq(a_c)\}^2}  +\frac{2}{4t^2|\partial_s\delta(s_c)\partial_aq(a_c)|} \\
    &=\frac{1}{4t^2} \left(\frac{1}{|\partial_s\delta(s_c)|} + \frac{1}{|\partial_aq(a_c)|}\right)^2.
\end{split}
\end{equation}

\section{The selection of POVM and CFI}\label{cal_CFI}
In App.~\ref{Holevo}, the tight bound at the crossing point is derived, then we find the POVM operators that realize that bound.
We construct the optimal measurement for each parameter $s$ and $a$ with the eigenvectors of the symmetric logarithmic derivatives (SLD) for the pure state defined as \cite{Pezze14}
\begin{equation}
\begin{split}
    \hat{L}_s = 2(\ket{\psi_f}\bra{\partial_s\psi_f} + \ket{\partial_s \psi_f} \bra{\psi_f}),\\
    \hat{L}_a = 2(\ket{\psi_f}\bra{\partial_a\psi_f} + \ket{\partial_a \psi_f} \bra{\psi_f}).
\end{split}
\end{equation}
At the crossing point, we have $\ket{\psi_f} = \ket{\psi_i} = (\ket{0} + i\ket{1})/\sqrt{2}$,  $\ket{\partial_s\psi_f} = -it(\ket{0} - i\ket{1})/\sqrt{2}$, and $\ket{\partial_a\psi_f} = t(\ket{0} - i\ket{1})/\sqrt{2}$.
Then the matrix forms of $\hat{L}_s$ and $\hat{L}_a$ are
\begin{equation}
\begin{split}
    \hat{L}_s = -t\begin{pmatrix}
        0 & 2\\
        2 & 0
    \end{pmatrix}\\
    \hat{L}_a = t\begin{pmatrix}
        2 & 0\\
        0 & -2
    \end{pmatrix}\\
\end{split}
\end{equation}
The eigenvectors of $\hat{L}_s$ are $\ket{+}$ and $\ket{-}$, and the eigenvectors of $\hat{L}_a$ are $\ket{0}$ and $\ket{1}$.
Therefore, the optimal POVMs for single parameter estimation for $s$ and $a$ are $\mathcal{M}_s=\{\ket{+}\bra{+}, \ket{-}\bra{-}\}$ and $\mathcal{M}_a=\{\ket{0}\bra{0}, \ket{1}\bra{1}\}$

To tackle the two-parameter estimation, we use these POVMs $\mathcal{M} = p\mathcal{M}_a + (1-p)\mathcal{M}_s = \{\ket{u_{m}} \bra{u_m}\}^4_{m=1}$ with $\ket{u_1} = \sqrt{p} \ket{0}$, $\ket{u_2} = \sqrt{p}  \ket{1}$, $\ket{u_3} = \sqrt{1-p}  \ket{+}$, and $\ket{u_4} =\sqrt{1-p} \ket{-}$.
This POVM corresponds to a measurement process in which a measurement is performed using the POVM $\mathcal{M}_a$ with a classical probability $p$ and $\mathcal{M}_s$ with a probability $(1 - p)$.
While it may appear arbitrarily chosen, we will demonstrate that adjusting the probability $p$, the Classical Cramér–Rao bound from CFI with these POVM matches the Holevo bound.
In order to derive the CFI, we first have
\begin{equation}
\begin{split}
    P(0 | s,a) &= \mathrm{Tr}[\ket{\psi_f}\bra{\psi_f} \ket{u_1} \bra{u_1}] = |\braket{u_1}{\psi_f}|^2  \\
    & = \frac{p}{2}\{(r_I + r_z)^2 + r_x^2 \} = \frac{p}{2}(1+ \frac{q}{\Omega} \sin{2 \Omega t}), \\
    P(1 | s,a) &= |\braket{u_2}{\psi_f}|^2= \frac{p}{2}(1 - \frac{q}{\Omega} \sin{2 \Omega t}), \\
    P(+ | s,a) &= |\braket{u_3}{\psi_f}|^2 = \frac{1-p}{2}(1 - \frac{\delta}{\Omega} \sin{2 \Omega t}), \\
    P(- | s,a) &= |\braket{u_4}{\psi_f}|^2 = \frac{1-p}{2}(1 + \frac{\delta}{\Omega} \sin{2 \Omega t}).
\end{split}
\end{equation}
Then,
\begin{equation}
\begin{split}
    \partial_s p(0 | s,a)  &= - \partial_s p(1 | s,a)  \\
    &= \frac{p\delta q}{2\Omega^3}(2\Omega t \cos{2 \Omega t}-\sin{2 \Omega t})\partial_s\delta,\\
    \partial_a p(0 | s,a) &= - \partial_a p(1 | s,a)\\
    & = \frac{p}{2}(\frac{1}{\Omega} \sin{2 \Omega t} - \frac{q^2}{\Omega^3} \sin{2 \Omega t} + \frac{ 2q^2t}{\Omega^2} \cos{2\Omega t}) \partial_a q \\
    &=\frac{p}{\Omega^3}(\delta^2 \sin{2 \Omega t} + 2q^2\Omega t \cos{2 \Omega t})\partial_aq,\\
    \partial_s p(+ | s,a)  &= - \partial_s p(- | s,a)\\
    &=-\frac{1-p}{\Omega^3}(q^2 \sin{2 \Omega t} + 2\delta^2\Omega t \cos{2 \Omega t})\partial_s\delta,\\
    \partial_a p(+ | s,a) &= - \partial_a p(- | s,a),  \\
    &=\frac{(1-p)\delta q}{2\Omega^3}(2\Omega t \cos{2 \Omega t}-\sin{2 \Omega t})\partial_aq.\\
\end{split}
\end{equation}

Near $s=s_c$, the CFI are
\begin{equation}
    \begin{split}
    \mathrm{C}_{ss} &= 4(1-p) t^2 {\rm sinc}^2(qt) \cos^2 (qt) (\partial_s\delta)^2\\
    \mathrm{C}_{aa} & = 4p t^2(\partial_a q)^2\\
    \mathrm{C}_{sa} & = 0
\end{split}
\end{equation}
Then, we have
\begin{equation}
    \mathrm{Tr}[\mathbf{C}^{-1}] = \frac{1}{4t^2} \left(\frac{1}{(1-p)(\partial_s \delta)^2} + \frac{1}{p(\partial_a q)^2}\right),
\end{equation}
at the crossing point.
For $p = |\partial_s \delta(a_c)|/\{|\partial_s \delta(s_c)| + |\partial_a q(a_c)|\}$, the classical bound is 
\begin{equation}
\begin{split}
    \mathrm{Tr}[\bm{\mathrm{C}}^{-1}] &= \frac{1}{4t^2} \left(\frac{|\partial_s \delta(s_c)| + |\partial_a q(a_c)|}{|\partial_aq(a_c)|\{\partial_s \delta(s_c)\}^2} + \frac{|\partial_s \delta(s_c)| + |\partial_a q(a_c)|}{|\partial_s \delta(s_c)|\{\partial_aq(a_c)\}^2}\right)\\
    &= \frac{1}{4t^2} \frac{(|\partial_s \delta(s_c)| + |\partial_aq(a_c)|)^2}{|\partial_aq(a_c) \partial_s \delta(s_c)|^2} \\
    &=\frac{1}{4t^2} \left(\frac{1}{|\partial_s\delta(s_c)|} + \frac{1}{|\partial_aq(a_c)|}\right)^2.
\end{split}
\end{equation}
This value is equal to the Holevo CR bound in Appendix~\ref{Holevo}, which indicates that the POVM we selected is optimal at the crossing point.

\section{Validity of the two-level approximation}\label{subspace}

\begin{figure}[h]
\centering
\includegraphics[width=\columnwidth]{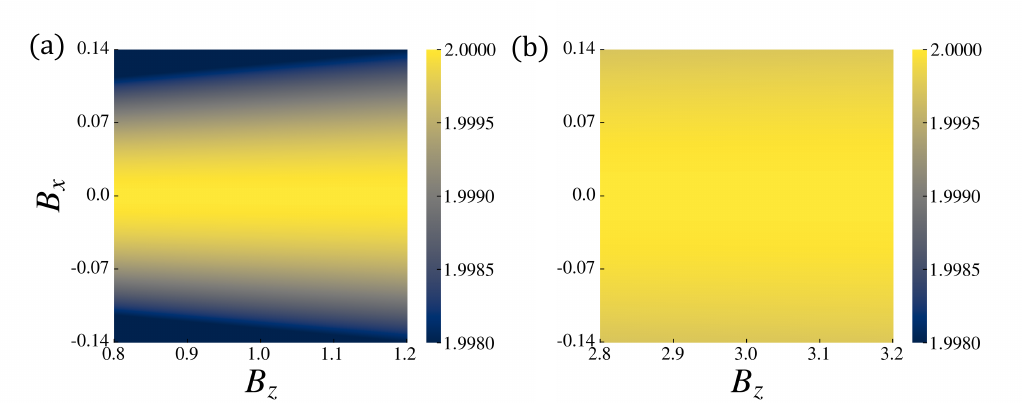}
\caption{Heatmaps of $S_2$, defined in Eq.~\eqref{Tr_norm}, for the two-spin Ising model with (a) $J=1$ and (b) $J=10$. Larger values of $S_2$ indicate better agreement between the two subspaces and, hence, a more accurate two-level approximation.}
\label{fig:Tr_norm}
\end{figure}

In this section, we quantify the accuracy of the two-level approximation in the two-spin Ising model.
For this purpose, we introduce a measure of the overlap between two $m$-dimensional subspaces of an $n$-dimensional Hilbert space.
Let ${\ket{a_i}}*{i=1}^m$ and ${\ket{b_i}}*{i=1}^m$ denote orthonormal bases of the two subspaces.
Their overlap is quantified by
\begin{equation}
S_2 = \sum_{i,j} |\braket{a_i}{b_j}|^2.
\label{Tr_norm}
\end{equation}
Equivalently, $S_2=\mathrm{Tr}(P_A P_B)$, where $P_A$ and $P_B$ are the projectors onto the two subspaces.
Thus, $0\leq S_2\leq m$, with $S_2=m$ when the two subspaces coincide and $S_2=0$ when they are orthogonal.

For the two-spin Ising model, we take ${\ket{a_i}}={\ket{v_1},\ket{v_2}}$ and ${\ket{b_i}}={\ket{\lambda_1},\ket{\lambda_2}}$, where $\ket{\lambda_i}$ are the eigenvectors corresponding to the eigenvalues $\lambda_i$ of the two-spin Ising Hamiltonian.
Figure~\ref{fig:Tr_norm} shows the resulting heatmaps of $S_2$.
Compared with $J=1$, the $J=10$ case exhibits a broader region with $S_2$ close to its maximum value of $2$, indicating a more accurate two-level approximation.
Note that $S_2$ always reaches its maximum value of $2$ at $B_x=0$, as can also be verified analytically from Eq.~\eqref{Ising_Hamiltonian_2}.

\bibliography{ref}

\end{document}